\documentclass[10pt]{article}

\usepackage[utf8]{inputenc}
\usepackage[T1]{fontenc}
\usepackage[margin=0.9in]{geometry}
\usepackage{url}
\usepackage{booktabs}
\usepackage{amsmath}
\usepackage{natbib}
\usepackage{enumitem}
\usepackage{titlesec}
\usepackage[hidelinks]{hyperref}

\setlist{noitemsep, topsep=1pt, partopsep=0pt}
\titlespacing*{\section}{0pt}{8pt plus 2pt minus 2pt}{4pt plus 1pt}
\titlespacing*{\subsection}{0pt}{6pt plus 2pt minus 2pt}{3pt plus 1pt}
\hypersetup{
  pdftitle={The Model Proposes, the Code Disposes: A Pre-Registered Ablation of a Verifier-and-Acceptance Stage in an LLM-Orchestrated Offensive-Security Agent},
  pdfauthor={Theodoros Moutesidis},
  pdfsubject={Ablation study of verifier and deterministic acceptance controls in an LLM-orchestrated offensive-security agent},
  pdfkeywords={LLM agents, offensive security, penetration testing, ablation study, verification, deterministic controls}
}

\title{\textbf{The Model Proposes, the Code Disposes}\\
\large A Pre-Registered Ablation of a Verifier-and-Acceptance Stage\\
in an LLM-Orchestrated Offensive-Security Agent}

\author{Theodoros Moutesidis \\ \emph{Independent Researcher}}

\date{}

\begin{document}
\maketitle

\begin{abstract}
We evaluate whether a verifier-and-acceptance stage -- a model verifier
whose verdicts are enforced by deterministic code -- changes what an
LLM-driven offensive-security agent reports. We report a 15-run exploratory
pilot, a pre-registered 20-run confirmatory ablation, and a pre-registered
$2\times2$ factorial study with 40 runs across two deliberately vulnerable
lab targets. In the confirmatory study, removing the stage eliminated
pre-report suppression (median 2 versus 0 findings per run; exact one-sided
$p=0.00003$) and reduced model-blinded shipped precision (median 0.471 versus
0.353; $p=0.0087$). Recall against a frozen but incomplete ground-truth list
did not differ significantly (two-sided $p=0.158$; equivalence was not
established). The factorial study attributed suppression to the model
verifier (Holm-adjusted $p=0.004$); deterministic acceptance rules alone
suppressed no false positives, and no interaction was detected ($p=0.72$).
The full design retained 93.8\% of model-adjudicated true candidates but did
not meet its pre-registered non-inferiority criterion because the lower
one-sided 95\% bound was 0.875, below the 0.90 floor. Across the confirmatory
and factorial studies, an instrumented
canary recorded zero contacts in 60 of 60 runs, with incidental external
contacts disclosed separately. Independent human adjudication of the
retained blind packets is pending, so precision and sensitivity endpoints
are supporting rather than final evidence. Six audit-trail failures,
including one in the evaluation tooling, are also disclosed. The results
support a narrow conclusion: the verifier changes what the system ships,
while deterministic code supplies enforcement and auditability; they do not
establish superiority to other agents or generalization beyond lab targets.
\end{abstract}

\section{Introduction}\label{sec:intro}

An offensive-security agent built on a large language model is, at its
core, a proposal generator. It reads a target, forms a hypothesis, and asks
to act on it. The hypothesis can be excellent or worthless, and the same
prompt can produce either on consecutive runs. Treating that variance as a
defect to be prompted away has not worked well in practice; treating it as
a fixed property of the model, and building deterministic infrastructure
around it, is the position this paper starts from. The model stays
probabilistic. The decisions about what it may do, what a claim must cite
before it is believed, how severity can move, what targets are in bounds,
and what gets reported when nothing was found, do not have to be. One
definition up front, because the title invites a stronger reading than the
data supports: ``disposes'' means that deterministic code owns
\emph{authority, acceptance mechanics, and accounting} -- which actions may
execute, which policy a verdict is applied under, and what the record says
happened. It does not mean code determines whether a vulnerability is true.
Truth assessment in this system is model work, and the factorial study in
Sec.~\ref{sec:factorial} measures exactly that division: the verification
effect belongs to a second model, while the deterministic layer's measured
contribution is enforcement and audit, not filtering. That
separation is the design thesis of the practitioner handbook this study is
drawn from, and it is not a new idea by itself -- deterministic gates around
probabilistic planners are a known pattern in agent engineering. What is
rarer is a controlled measurement of one such gate's effect, reported with
its own instrument failures left in.

Self-evaluation of an agent framework is usually unconvincing for a
structural reason, not a moral one: one team designs the system, drives the
runs, defines the outcome, and grades the transcript. Every one of those
roles has a natural incentive to round an ambiguous result toward the story
already believed, and a reader cannot see where the rounding happened. A
three-run pilot showing ``every treatment run above every control run''
sounds like a finding; it is also exactly what a coin flip produces once
every ten pairwise comparisons at that sample size, and a paper that never
runs that arithmetic will not tell a reader which one it got. Our response
is to publish the part of this process normally cut first: we ran an
adversarial review of our own pilot, disclose that the review was itself
LLM-performed, and report what it invalidated before reporting what
survived. We then pre-registered a second, larger study of the one claim
the review left standing, froze the design before the first run, and
treated a bug found in our own grading instrument as a result to report
rather than an embarrassment to fix quietly.

This paper's contributions:

\begin{enumerate}
\item A concept-level description of six deterministic control layers
  around an LLM-orchestrated offensive-security agent (Sec.~\ref{sec:system}), without
  proprietary implementation detail.
\item A disclosed 15-run exploratory pilot and the adversarial review that
  invalidated its headline statistics while leaving a defect-discovery
  record intact (Sec.~\ref{sec:pilot}).
\item A pre-registered, cold-state, 20-run confirmatory study of one
  control layer -- the verifier-and-acceptance stage -- with exact
  non-parametric tests, a frozen exclusion rule, and a complete deviations
  log (Sec.~\ref{sec:study}).
\item A pre-registered $2\times2$ factorial bridge study (40 cold-state
  runs, two lab targets) that separates the stage's model verifier from its
  deterministic acceptance rules, attributes the measured effect to the
  verifier, reports a null interaction at equal billing, and quantifies the
  design's sensitivity cost against a floor it did not clear
  (Sec.~\ref{sec:factorial}).
\item A ``failure museum'' of five reproducible control-mechanism defects
  plus a sixth, self-caught defect in the evaluation tooling, each traced
  to two individually correct rules composing incorrectly (Sec.~\ref{sec:museum}).
\item An explicit safety and containment ledger, and a limitations section
  written more generously than an external reviewer would require
  (Sec.~\ref{sec:ledger}--\ref{sec:limits}).
\end{enumerate}

We make no claim that this framework outperforms any other offensive-agent
system, no time or cost claim, and no claim of clean separation between
arms. The headline result is a tested median difference on one public lab
target, reported with its overlap stated beside it.

\section{System Context}\label{sec:system}

The system under test is an LLM-orchestrated security-testing framework.
An orchestrating model plans a test of a declared target and
proposes actions; deterministic layers sit between that model and any
effect in the world. We describe these at the level needed to interpret
the study, not at build-manual depth.

\textbf{One write path for findings and evidence.} Every finding, scope
decision, and recorded probe result the run persists as a result flows
through a single code-owned writer; every other layer assumes this has
already run. This is a guarantee about persisted findings and evidence, not
a sandboxing claim: the orchestrator can run shell commands and works out of
a shared per-run scratchpad, and both studies logged the failure mode that
gap predicts -- one bare-curl slip in the pilot that bypassed the write path,
self-caught and redone through it (Sec.~\ref{sec:pilot}), and one stale-scratchpad
near-miss in the confirmatory study, caught and corrected before it touched
a real record (Sec.~\ref{sec:study-deviations}) -- rather than a channel this design closes off
outright.

\textbf{Evidence binding.} A claim about something already observed must
cite a captured artifact: a request/response pair, a quoted string. A
matching quote establishes the citation is genuine, not that the
conclusion is true. A proposal about a future action is a different kind of
statement and passes a different gate -- authorization, scope, budget --
before it runs at all.

\textbf{A downward governor.} A deterministic severity policy can lower a
reported severity, or mark a finding false, and cannot raise either. This
bounds how much damage an overclaiming model can do to a report's
credibility. It does not, alone, protect against under-claiming; Sec.~\ref{sec:museum}
shows what happens when it composes badly with a neighboring rule.

\textbf{Verifier-and-acceptance stage.} A narrow review step -- a model
verifier plus deterministic handling of its verdicts -- evaluates a proposed
finding against its bound evidence before the finding may ship, and is the
only layer permitted to raise a governor-capped severity, and only against a
verbatim quote. This is the layer ablated below.

\textbf{Scope layers.} Authorization is a policy checked before each
outgoing action, not stated once in a prompt and trusted to survive the
rest of the context window. The design deliberately stacks two independent
layers -- a standing rule and a per-run declaration -- so losing one does
not remove containment; Sec.~\ref{sec:pilot} reports what happened when one layer was
removed and the other held.

\textbf{Honest accounting.} The deliverable records what the run did not
do -- targets skipped, actions refused, baselines left unmeasured -- next to
what it found, so a clean scan and a scan that could not reach anything are
not rendered identically.

None of these layers make the model good at finding vulnerabilities. Both
arms of the confirmatory study repeatedly found several of the same core
vulnerability classes, with substantial per-run variation (Sec.~\ref{sec:study-results}). What
these layers are argued to do is make the model's output governable,
auditable, and safe to run unattended, and the empirical question here is
how much one of them -- the verifier-and-acceptance stage -- actually
contributes to that, measured rather than argued.

\section{The Exploratory Pilot, and Its Demolition}\label{sec:pilot}

Before pre-registering anything, we ran a five-arm exploratory matrix: a
full-rules arm and four arms each removing one control (the
verifier-and-acceptance stage, a memory subsystem, a scope declaration, a
grounding-critic gate), three runs per arm, against the OWASP Juice Shop
training
application, with state carried warm across the three runs inside each arm.
An adjudication rubric, written after inspecting the runs, scored the
shipped findings for precision. This pilot's own write-up called its
headline comparison ``clean.''

It was not. We subjected the pilot to an adversarial review -- itself
LLM-performed and disclosed as such -- instructed to break the result, not
polish it. It invalidated the pilot's central claims on four grounds:

\begin{itemize}
\item The ``analyst-level'' adjudication was an eight-branch keyword
  classifier tuned on the exact titles this 15-run corpus produced, ending
  in a catch-all defaulting to false. Cross-referenced against the
  framework's own severity-governor log, the identical finding title
  appeared on both sides of a contradiction -- capped as real by the
  governor, branded false by the classifier -- in 7 of 15 runs, with no
  tiebreaker and no rationale recorded for either side.
\item ``The verifier effect survives blinding, cleanly'' was confounded
  with a re-packaging change in the adjudication pipeline, not with the
  arms themselves. A corrected packet-assembly pass moved the full-rules
  arm from worst-scoring in the entire matrix to best, while the
  no-verifier arm's per-run verdicts did not move at all between packet
  versions. The wanted direction was real; the instrument that produced it
  was not yet trustworthy when it did.
\item ``Every full-rules run scored above every no-verifier run'' is
  exactly what a null distribution predicts once per ten pairwise arm
  comparisons at three runs per arm -- the review supplied the permutation
  arithmetic the original write-up never ran, and the observed count (one
  such separation) matched the count expected under pure noise.
\item The grounding-critic arm's own prompt told the orchestrator, inside
  the instruction, that it was reproducing ``a controlled arm'' of an
  experiment -- a textbook demand characteristic invalidating any claim
  built on that arm's counter-intuitive result.
\end{itemize}

A fifth correction: the pilot's safety count, ``canary contacted zero times
in 15 of 15 runs,'' silently dropped an unmeasured run and rounded it to
zero. The correct count was 0 of 14 measured, with the fifteenth -- the
coldest, first-executed run of the flagship arm -- simply not instrumented
for that measurement. We report it as 0/14 measured, 1 unmeasured.

What survived without needing any adjudication: all 15 runs completed and
finalized honestly; the catalogued, off-the-shelf scanning tools bundled
with the framework never independently confirmed a single exploit that
hand-driven, evidence-bound probing found in the same run; bare access to
the target was structurally unavailable from the orchestrator's shell, and
the single bare-curl slip that did occur was self-caught, discarded, and
redone through the controlled write path rather than shipped from outside
it; the undeclared canary target was untouched in every measured run,
including three runs with no per-run scope declaration at all; and five
reproducible, code-level mechanism defects were found by the runs' own
audit trail (Sec.~\ref{sec:museum}). None of that required trusting the contested
precision numbers.

We treat this pilot as a defect-generating exercise, not evidence for or
against any control layer's effect on precision or recall. Its only
downstream contribution is the to-do list that Sec.~\ref{sec:study}'s pre-registration
was written to close: drop the adjudication rubric that scores its own
answers, run cold instead of warm, measure one variable instead of four
confounded ones, fix the statistic before the data exists, and do not tell
the model under test it is in an experiment.

\section{Confirmatory Study}\label{sec:study}

We pre-registered a two-arm study of the one control the pilot's review
left standing as worth a real follow-up: the verifier-and-acceptance stage
(Sec.~\ref{sec:system}). The study ablates that stage as a package -- the model verifier
together with the deterministic handling of its verdicts -- and does not
isolate the deterministic component alone. The design was frozen in a dated
document before run 1; every field below is taken from it, and every
departure during execution is logged as a deviation, not silently absorbed.

\subsection{Design}

\textbf{Arms.} FULL (stage active) versus NO-VERIFY (stage skipped), $n=10$
runs per arm, 20 total.

\textbf{Target and state.} OWASP Juice Shop \citep{owaspjuiceshop}, a fresh container per run,
cold state throughout: no memory or configuration carried over between
runs, removing the warm-state non-independence of the pilot's runs 2 and 3
within each arm.

\textbf{Order.} All 20 runs interleaved in a seeded random permutation
fixed and recorded before run 1.

\textbf{Models and prompts.} Each run used one Claude Sonnet 5 orchestrator
through Claude Code (high effort, standard speed) under a fixed harness
revision. The host did not expose a temperature setting or deterministic
sampling seed. The two arm
prompts are byte-identical except one mechanical configuration sentence in
NO-VERIFY, disclosing the toggle through the same channel the production
system uses. Neither prompt names ``experiment,'' ``ablation,'' or
``measure,'' or refers to any prior run or expected outcome. We disclose,
rather than dismiss, the resulting limitation: the model can in principle
read its own configuration sentence and behave differently knowing which
arm it is in. A further disclosure, detailed with its sensitivity check in
Sec.~\ref{sec:study-deviations}: one arm-symmetric safety amendment mid-study means the 20 prompts
were not byte-identical across the whole study -- there are two prompt
periods, byte-identical between arms only within each.

\textbf{Exclusion rule, written before run 1.} A run is excluded only for
infrastructure failure -- scoring or dashboard service unreachable at
launch, or the orchestrator terminating before the first pipeline stage
returns a run identifier -- logged and replaced with another run of the
same arm at the schedule's end. Zero runs were excluded; all 20 scheduled
runs are analyzed as run.

\textbf{Endpoints, fixed in advance.} P1, pre-report suppression (primary,
mechanical, adjudication-free): findings flagged false before the report
ships. P2, severity-integrity events (primary, no pre-specified direction,
but a two-sided exact test was itself pre-specified): shipped findings
whose evidence supported a higher band than they shipped at. P3,
shipped-finding count (secondary). P4, blinded shipped precision
(secondary): a fresh model adjudicator, working from a label-stripped
packet assembled from every evidence store the run produced, under a frozen
scoring guide. The adjudicator was a fresh session from the same model
family; its exact model snapshot was not recorded in the verdict artifact,
which limits reproducibility. The same packet and key format also produce a
human-pass bundle, designated the registered independent check for this endpoint. P5, recall
against a 20-entry curated ground-truth list frozen before the study, with
no pre-registered direction and no non-inferiority margin. Canary contacts,
wall time, and tool-call counts were tertiary, with no hypothesis attached.

\textbf{Hypotheses and tests, fixed in advance.} H1: P1 higher under FULL
than NO-VERIFY, exact one-sided Mann--Whitney U \citep{mannwhitney1947},
$\alpha=0.05$. H2: P2,
direction not pre-specified; a two-sided exact test, reported descriptively
if ties dominate. H3: P4 higher under FULL, same one-sided test, reported
as supporting evidence only until a human blind pass exists. P5 carried no
directional hypothesis and no equivalence claim was pre-registered for it.
No other comparison is headline-eligible.

\subsection{Results}\label{sec:study-results}

\begin{table}[t]
\centering
\caption{Confirmatory study results (10+10 cold runs, seeded interleaved order, frozen pre-registration, zero exclusions). Values for P1, P3--P5 are per-run medians; P2 is a total across the 10 runs of each arm; the canary row is a total contact count. All tests were pre-specified before run 1.}
\label{tab:confirmatory-results}
\footnotesize
\begin{tabular}{@{}p{0.29\linewidth}rrp{0.25\linewidth}p{0.16\linewidth}@{}}
\toprule
Endpoint & FULL & NO-VERIFY & Test & Outcome \\
\midrule
P1 pre-report suppression (primary) & 2.0 & 0.0 & one-sided exact MWU, $U=98/100$, $p=0.00003$ & H1 confirmed \\
P2 severity-integrity events (total) & 17 & 15 & two-sided exact MWU (as pre-registered), $p=0.854$ & no detected difference \\
P3 shipped findings & 16.5 & 17.5 & two-sided exact MWU, $p=0.054$ & marginal \\
P4 blinded shipped precision (secondary) & 0.471 & 0.353 & one-sided exact MWU, $U=81/100$, $p=0.0087$ & preliminary support\textsuperscript{\ensuremath{\dagger}} \\
P5 curated ground-truth recall\textsuperscript{\ensuremath{\ddagger}} & 0.25 & 0.20 & two-sided exact MWU, $p=0.158$ & not significant \\
Canary contacts (tertiary, total)\textsuperscript{\S} & 0/10 & 0/10 & -- & 0/20 measured \\
\bottomrule
\end{tabular}

\vspace{2pt}
\raggedright\scriptsize
\textsuperscript{\ensuremath{\dagger}}Model-blinded adjudication: H3 has preliminary support -- the registered human blind pass on the retained, label-stripped packet. P4 distributions overlap at their edges (minimum FULL run = 0.294, maximum NO-VERIFY run = 0.476): the reported effect is a tested median difference, not a clean separation. Per-run P1 values, FULL: 2, 2, 8, 2, 2, 1, 10, 1, 2, 8 (every run $\geq 1$); NO-VERIFY: 0, 0, 0, 1, 1, 0, 0, 0, 0, 0 (the two 1s trace to a write-time governor action, not to any verifier).\\
\textsuperscript{\ensuremath{\ddagger}}P5 carried no pre-registered direction; this two-sided test is not a claim of equivalent recall (no non-inferiority margin was set). The frozen 20-entry ground-truth list also double-counts one condition -- the login SQL-injection finding satisfies both an \texttt{sqli} and an \texttt{auth\_bypass} entry at the same endpoint, in all 20 runs. Deduplicated to 19 distinct conditions: FULL median 0.211, NO-VERIFY median 0.158, two-sided $p=0.158$; the inflation is uniform across arms (roughly five percentage points), so the values above are overstated by construction, not by arm.\\
\textsuperscript{\S}The instrumented on-network canary was contacted 0 times in 20 of 20 measured runs, alongside a disclosed crawler redirect-follow to a public code-hosting page in most runs and one external DNS lookup per run (Sec.~\ref{sec:ledger}); this is a canary-contact count, not a general containment guarantee.
\end{table}

Table~\ref{tab:confirmatory-results} reports the pre-registered endpoints
exactly as specified. Removing the verifier-and-acceptance stage eliminated
pre-report suppression (median 2 under FULL, 0 under NO-VERIFY, exact
one-sided $p=0.00003$; H1 confirmed) and reduced label-blinded shipped
precision (median 0.471 under FULL, 0.353 under NO-VERIFY, $p=0.0087$; H3
has preliminary support from model-blinded adjudication, with the registered
human blind pass pending). Recall against the curated
ground-truth list showed no statistically significant difference (0.25
vs.\ 0.20, two-sided $p=0.158$); P5 carried no pre-registered direction,
and the study was not designed to establish equivalent recall -- no
non-inferiority margin was set, so this is an absence of detected
difference, not evidence of equivalence. The frozen 20-entry list also
double-counts one condition: the login SQL-injection finding satisfies
both an \texttt{sqli} and an \texttt{auth\_bypass} ground-truth entry at
the same endpoint, in all 20 runs. Deduplicated to 19 distinct conditions,
recall is 0.211 under FULL versus 0.158 under NO-VERIFY (two-sided
$p=0.158$), with the inflation uniform across arms at roughly five
percentage points, so it changes the absolute recall values above but not
which arm they favor. Severity-integrity events showed no detected
difference across arms (17 vs.\ 15 total; two-sided exact MWU, $p=0.854$,
as pre-registered), consistent with severity-band write-time governance
operating independently of the verifier. The instrumented on-network
canary was contacted zero times in all 20 measured runs, alongside the
disclosed crawler redirect-follow and per-run DNS lookup detailed in
Sec.~\ref{sec:ledger}.

We state the finding at its true strength rather than its most dramatic
phrasing. This is a tested median difference, not a clean separation: the
minimum precision score among FULL runs (0.294) is below the maximum score
among NO-VERIFY runs (0.476). Both arms repeatedly found several of the
same core vulnerability classes, with substantial per-run variation: a
SQL-injection authentication bypass at the login endpoint appeared in
every run's findings; a UNION-based data extraction and an insecure direct
object reference on the shopping-basket endpoint appeared in most runs; a
mass-assignment privilege escalation shipped in 6 of 10 FULL runs and 3 of
10 NO-VERIFY runs. The stage's clearest measured effect is on what
shipped, not on what the orchestrating model was capable of finding.

Two qualitative patterns appeared in both directions and are worth stating
plainly. NO-VERIFY runs twice identified false positives the orchestrator
could not itself remove without the stage, and one NO-VERIFY run lost a
genuine finding -- a null-byte path-traversal bypass -- to an unrelated
write-time governor rule misfiring on an embedded negative control, with no
repair path once the verifier was off. FULL runs' verifiers, conversely,
caught false positives including one orchestrator's own accidental
duplicate submission, and twice forced an orchestrator to produce a
verbatim quote before a capped severity was raised back to its supported
band. This stage is, in this system, both the principal noise filter and
the only severity-repair mechanism; removing it removes both at once.

\subsection{Deviations}\label{sec:study-deviations}

All six departures from the frozen design, as committed to by the
pre-registration, are reported here.

\begin{enumerate}
\item A residual redaction gap, found before run 4: a redaction routine
  masked structured credential fields but missed one alternate encoding, so
  a probe's raw capture carried lab-only authentication material into a
  raw-evidence field despite a correctly redacted summary.
  We did not patch mid-study -- the pre-registration pins the system
  revision, and the gap was arm-symmetric -- and fixed it after the study
  closed, scrubbing affected rows across all runs before any packet was
  shared.
\item A new late-stage regression-sweep behavior, introduced by the pinned
  harness revision, appeared identically in both arms: a uniform,
  non-differential change.
\item One NO-VERIFY run's probe layer briefly held an unbounded data
  extraction before the orchestrator re-probed with a bounded query on its
  own initiative; only the bounded result was stored.
\item After run 11, both arm prompts gained one identical sentence
  prohibiting injection-based writes capable of affecting more than one
  record. This followed two self-reported incidents -- one per arm -- where
  a model-authored filter matched and updated every row of a table instead
  of one, on the disposable, per-run target container rebuilt before every
  run. Blast radius was confined to that container; the amendment applied
  prospectively and identically to both arms from run 12 onward, preserving
  prompt symmetry \emph{between} the arms. Stated plainly, it does not
  preserve prompt symmetry \emph{across} the whole study: the 20 prompts
  were not byte-identical from run 1 to run 20, but fell into two periods
  (runs 1--11 and 12--20), byte-identical between arms only within each. A
  descriptive sensitivity check on the primary endpoint: the pre-report
  suppression direction holds in both periods (FULL median 2.0 pre-amendment
  versus 5.0 post; NO-VERIFY median 0 in both), so the amendment does not
  explain the primary effect, though it remains a disclosed time/prompt
  confound.
\item One run showed a near-miss where a stale scratchpad file from a
  prior run briefly caused a write to land against an unrelated,
  already-completed run; caught and corrected immediately, no data
  recorded against the wrong run. Per-run scratchpad isolation remains an
  open infrastructure item.
\item One run raised a suspicion of unintended network egress from a
  cloud-misconfiguration check; a post-study code review resolved it. Such
  checks are reference-gated on names appearing in the target's own pages,
  none of which this target contained, so no provider contact occurred in
  any run. The one unconditional external touch in every run is a single
  DNS lookup via the system resolver, flagged as an item to gate more
  tightly for non-public targets.
\end{enumerate}

\section{Factorial Bridge Study}\label{sec:factorial}

The confirmatory study ablated a package: a model verifier plus the
deterministic handling of its verdicts, switched off together, because that
was the only off-switch run. A skeptical reading of
Sec.~\ref{sec:study-results} therefore cannot say which component earned the
result. The bridge study answers that with a pre-registered $2\times2$
factorial: factor V (the model verifier, stage on/off) crossed with factor R
(deterministic acceptance -- the severity governor and the
accept/refuse/raise verdict policy). Four arms, five cold-state runs per arm
per target, two targets (OWASP Juice Shop for continuity, plus VAmPI
\citep{vampi}, a deliberately vulnerable REST API), 40 runs in one cohort with zero
exclusions. The design, endpoints, hypotheses, test statistics, and
multiple-comparison correction were frozen before run one. Factor V is
communicated by the same single configuration sentence the confirmatory
study used (a disclosed limitation); factor R is environment-enforced at the
scanner and invisible to the orchestrator's instructions, and every verdict's
audit record carries the acceptance policy actually in force, so the
assignment is verifiable per run after the fact. The orchestrators again
used Claude Sonnet 5; the blinded factorial adjudicator used Claude Opus, a
distinct model in the same family.

\subsection{Results}\label{sec:factorial-results}

\begin{table}[t]
\centering
\caption{Factorial bridge study: per-arm outcomes over 40 runs (5 per cell, two targets). J1 medians are shown for the informative target (Juice Shop); the second target sits at the precision ceiling in every arm ($\approx 1.0$) and carries no discrimination power. J2 is pooled per arm over the blinded adjudication of all 561 candidates, including the 34 rejected ones. Model-blinded throughout; human blind pass pending.}
\label{tab:factorial-results}
\small
\begin{tabular}{@{}lrrrr@{}}
\toprule
 & A0 & A1 & A2 & A3 \\
 & (neither) & (rules only) & (verifier only) & (both) \\
\midrule
J1 blinded precision (median, Juice) & 0.400 & 0.312 & 0.444 & 0.500 \\
J2 sensitivity (TP shipped / TP cand.) & 0.962 & 1.000 & 0.989 & 0.938 \\
J2 specificity (FP caught / FP cand.) & 0.017 & 0.000 & 0.175 & 0.226 \\
Suppressions: verifier verdicts & 0 & 0 & 11 & 18 \\
Suppressions: governor rules & 0 & 0 & 0 & 0 \\
Suppressions: base write-time heuristics & 4 & 0 & 0 & 1 \\
GT recall, median (Juice / VAmPI) & \multicolumn{4}{c}{0.158 / 0.571 -- identical across all arms} \\
Canary contacts & \multicolumn{4}{c}{0 in 40 of 40 runs} \\
\bottomrule
\end{tabular}
\end{table}

Table~\ref{tab:factorial-results} reports the per-arm outcomes. The two
pre-registered primaries, Holm-corrected as a family \citep{holm1979}: H1, the verifier's
main effect on pre-report suppression with no deterministic acceptance
behind it (verifier-only vs.\ neither), confirmed -- stratified exact
permutation over per-target Mann--Whitney statistics, one-sided $p = 0.002$,
Holm-adjusted $p = 0.004$. H2, the verifier$\times$rules interaction on
label-blinded shipped precision, null -- interaction contrast
$T = +0.04$, two-sided seeded Monte Carlo permutation $p = 0.72$ (the
pre-registered exact enumeration is computationally infeasible for four
groups; the substitution is disclosed in the deviations log). The
pre-registration committed to publishing this null at equal billing, and it
is not a technicality concealing a near-miss: on the informative target the
rules-only arm's blinded precision median (0.312) sits below the raw arm's
(0.400), while both verifier arms sit above it (0.444, 0.500).

The per-layer suppression audit is the study's bluntest row: across all 40
runs, the deterministic governor's mark-false-positive action fired zero
times. Every measured act of false-positive suppression above the always-on
write-time heuristics came from verifier verdicts (29 in the two verifier
arms). The same audit cuts against the base heuristics: they false-flagged
five genuine findings across the study (a pattern heuristic misreading a
finding's own prose), and only verifier-equipped runs could repair the
damage -- three of those findings were restored by verdicts in the verifier
arms, while the raw arms shipped their misfires as dead findings. Rules
without a verifier cannot correct their own errors; that asymmetry, not any
suppression the rules perform themselves, is the deterministic layer's
measured relationship to report quality here.

The model-adjudicated estimate of the full design's cost is a result, not a
footnote. H4
pre-registered a non-inferiority bar -- the full design keeps at least 90\%
of adjudicated-true findings, demonstrated at 95\% confidence -- and the
data did not clear it: sensitivity 0.938 (76 of 81), lower one-sided 95\%
Clopper--Pearson bound 0.875 \citep{clopperpearson1934}. Part of the shortfall has a named, fixable
mechanism: one run's genuinely executed account-takeover finding was
rejected because evidence redaction masked the identity-bearing header past
the point where the verifier's required verbatim quote could exist. H5's
recall equivalence, this time with a pre-registered $\pm$1-item margin the
confirmatory study lacked, held exactly: median deduplicated ground-truth
recall identical in all four arms on both targets (0.158 Juice Shop, 0.571
VAmPI). And the confirmatory comparison replicated out of sample inside the
factorial: full design versus rules-only blinded precision, 0.500
vs.\ 0.312, same direction as the confirmatory 0.471 vs.\ 0.353, in
different runs under a different randomization.

H3, the severity-integrity contrast between the shipped asymmetric verdict
policy and the raw policy, was null by absence of events: no verifier in any
arm attempted a severity raise without a verbatim quote, so the asymmetric
rule's raise gate remains tested by construction and unit test, not by field
data. The verifiers did apply severity \emph{downgrades} as given in the raw
arms -- audit-trail material, but identically legal under both policies.

\subsection{Deviations and scope integrity}\label{sec:factorial-deviations}

The full log accompanies the artifacts; the entries that bear on
interpretation: the crawler's known redirect-follow contacted the Juice Shop
target's own public code-hosting page in 11 of 20 Juice runs -- the behavior
proved engine-dependent (it occurred under both browser engines and
demonstrably did not occur in several runs whose traffic captures were
checked), so it is flagged per run rather than assumed constant, and it is
arm-symmetric; one run read the target application's source from inside the
in-scope container (read-only, all claims still verified by live requests;
flagged, and its exclusion changes no conclusion); one certificate
transparency lookup on a bare container name returned nothing; the
instrumented canary was contacted zero times in 40 of 40 runs; the blinded
adjudicator is a distinct model from the orchestrators but the same model
family (no non-Claude adjudicator was available), and the human blind pass
over the identical 561-item packet -- which includes the 34 rejected
findings, indistinguishable from shipped ones -- remains the registered
independent check for every model-adjudicated number in this section.

For the framework this paper describes, the attribution correction is the
finding: where earlier drafts let the deterministic layer share credit for
cleaner reports, the measured split assigns suppression and precision to the
model verifier, and re-scopes the deterministic acceptance layer's
defensible claims to severity governance, duplicate control, auditability of
verdicts, and a small additional specificity on top of the verifier (+0.05,
descriptive). If a builder can afford only one of the two components, these
model-adjudicated data favor prioritizing the verifier, subject to the
registered human check.

\section{Failure Museum}\label{sec:museum}

The audit-trail claim in Sec.~\ref{sec:system} is worth only what it actually catches.
This section reports six defects in the museum's own convention: what
happened, why the individual rules involved were each correct on their own
terms, and what changed. These are implementation-level findings; internal
identifiers and exploit-specific values are omitted, and we do
not claim any is now provably eliminated beyond the fix or disclosure
stated.

\textbf{1. The attachment that unredacted the finding.} A convenience path
that attached a probe's full raw request and response to a finding record,
for provenance, copied lab-only authentication material into a raw-evidence
field beside a correctly redacted summary field. The redaction routine
worked exactly as specified; the attachment convenience also worked exactly
as specified. Neither rule was wrong in isolation. The fix moves redaction
to the single write path, so every copy a record carries is masked, not
only the field an author inspects.

\textbf{2. The kill-switch that half-killed.} An environment flag
advertised as disabling a memory subsystem stopped that subsystem's writes
and its own construction. Two read endpoints, however, assembled prior-run
history directly from storage without consulting the flag, and one direct
record call fell back to an older writer predating the flag. Each
component, alone, did what its own contract promised. A study arm built on
this switch measured ``memory off'' while receiving the full recall of
every prior run. The fix extends the capability freeze to every read lane
and fallback, not only the constructor.

\textbf{3. Two correct rules, one wrong severity.} A proven, unauthenticated
administrative takeover, submitted with verbatim quoted proof in the
system's canonical structured shape, shipped at a middling severity instead
of critical. The evidence store's mirror field accepted only string-typed
evidence; the structured proof sat one field away, serialized and ignored,
so the severity grader graded an effectively empty input as thin, and the
downward governor correctly capped severity to what a thin grade allows.
Every rule -- hygiene, grader, governor, raise rule -- was byte-correct; the
composition was wrong. Reproduced across three configurations before the
fix: canonicalize evidence at the single write path, so grading no longer
depends on which field an author populated.

\textbf{4. Dedup ate the correction.} An author who noticed the
under-labeled finding above resubmitted it with the missing proof attached.
Deduplication matched on type, URL, and title, found an exact match, and
silently returned the existing record -- exactly its intended job for a
genuine repeat. The resubmission was not a repeat; it was a correction
carrying materially better evidence, and the absorption meant it never
reached governance. The audit log recorded a duplicate skipped, which reads
as healthy, not a repair refused. The fix teaches the collapsing rule to
distinguish ``the same claim again'' from ``the same claim, now with
proof,'' and lets the second kind through.

\textbf{5. Coverage that could not see the work.} A coverage percentage,
built after an earlier incident to stop a scan claiming credit it had not
earned, credited catalogued tool runs against a fixed URL list. Both
design choices were individually sound. What it could not see was the
hand-driven, evidence-bound probing channel that produced every confirmed
exploit here; its denominator also absorbed URLs from a page-scraping
component with a malformed origin string nothing could ever test. The same
metric family had previously reported near-total coverage on a run
executing no active tool, and here reported near-zero coverage on runs
extracting real data with bounded proof -- opposite failures, one root
cause: the denominator was not the work. Probe-technique coverage is now
counted; canonicalizing the baseline URL set remains open.

\textbf{6. The evaluation tool that graded its own blind spot.} An early
blinded-adjudication packet, built during the pilot, fed a fresh model
adjudicator only the human-readable evidence-summary field of each finding.
Findings whose proof lived in a raw-evidence field -- a mechanical
consequence of which pipeline stage produced them, not of whether the
claim was true -- were auto-scored false. This flipped the apparent
direction of a between-arm precision comparison in an early pass: the
full-rules arm briefly looked worst in the entire matrix. Reconciliation
against the framework's own governance log caught the discrepancy before
publication. Both the packet-building choice and the storage layout
separating a summary field from a raw-evidence field (Entry 1's redaction
boundary) were individually reasonable; composed, they silently inverted a
headline number. The corrected packet assembles from every evidence store a
finding touched, not one field. This is why Sec.~\ref{sec:study} reports its precision
endpoint as model-blinded with a human pass pending, not closed: an
instrument that failed once, silently, until an unrelated reconciliation
step caught it, does not get to grade its own rematch unsupervised. This
entry demonstrates, as much as any p-value here, the audit-trail claim of
Sec.~\ref{sec:system}: the same discipline -- bind every claim to a capture, reconcile
before shipping -- caught an error in the instrument measuring the system,
not only in the system itself.

\section{Safety and Containment Ledger}\label{sec:ledger}

All work reported here ran against two disposable, isolated, open-source
training applications, never a production or third-party system. We report
the full ledger rather than a summary that would flatter it.

\textbf{Canary.} 0 of 14 measured pilot runs contacted an instrumented
on-network canary target, with the pilot's fifteenth run correctly reported
as unmeasured, not zero (Sec.~\ref{sec:pilot}). The confirmatory study measured all 20
runs and recorded zero contacts in every one. Read together with the
incidental contact disclosed next, this is a canary-contact count, not a
general containment guarantee: it does not assert that scope containment
held unqualified in every run.

\textbf{Incidental network contact.} A crawling component's
redirect-follow behavior fetched an off-target code-hosting page in the
pilot and confirmatory study and in 11 of 20 factorial Juice Shop runs,
disclosed per run. The confirmatory study's only other external touch was one
unconditional DNS lookup per run; a suspected cloud-provider egress path
was investigated and resolved as never firing, since the checks that could
cause it are gated on names appearing in the target's own content, which
this target lacks.

\textbf{Destructive actions.} A destructive stacked-query step was
correctly deferred for operator approval during the pilot even under an
aggressive testing profile. The confirmatory study recorded two mass-write
incidents -- one per arm -- where a model-authored filter matched and
updated every row of a table instead of the one a proof-of-concept
requires. Both were self-reported, both confined to a disposable,
per-run-rebuilt container, and both led directly to the mid-study,
arm-symmetric prompt amendment in Sec.~\ref{sec:study-deviations}.

\textbf{Credential hygiene.} Lab credentials were not used outside the
disposable targets. The redaction gap in Sec.~\ref{sec:study-deviations} and Entry 1 of Sec.~\ref{sec:museum} let
lab-only authentication material reach a raw-evidence field in a small
number of runs; affected rows across both
studies were located and scrubbed after the confirmatory study closed,
before any adjudication packet was shared beyond the study team.

Every containment number here is a count over runs against two disposable
open-source lab targets: evidence about this system's behavior under these
conditions, not a general claim about containment against production
infrastructure.

\section{Limitations and Threats to Validity}\label{sec:limits}

We state these limitations explicitly because self-evaluation magnifies the
risks of shared assumptions, incomplete instrumentation, and selective
interpretation.

\textbf{Lab targets only.} The pilot and confirmatory studies ran
exclusively against one public, deliberately vulnerable single-page
application backed by a lightweight database; the factorial added a second
lab target, a small REST API that sits at the precision ceiling in every arm
and therefore discriminates nothing on that endpoint. Nothing here
generalizes to other architectures, stacks, or live engagements.

\textbf{Interaction power.} The factorial's null interaction rests on five
runs per cell; it is an absence of detected effect, not a demonstration of
no effect. The same caution applies in reverse to the rules-only arm's
below-raw precision median, which we report descriptively and decline to
test post hoc.

\textbf{Model-family circularity.} The orchestrators, the adjudicators
that scored their output, the defects discovered, and the fixes applied
were all produced within one model family and one operator lineage. Both
studies' blinded adjudicators are models from the same family as the
orchestrators they grade (the factorial's is a distinct model within it;
the confirmatory verdict artifact did not record the exact adjudicator
snapshot). A
human blind pass on each retained, label-stripped packet is the registered
independent check for every adjudicated endpoint and has not yet run.

\textbf{Configuration by instruction.} The arm condition is communicated to
the orchestrator as one sentence in its own configuration context -- the
same channel production uses -- so the model can in principle read which
arm it is in. We found no behavior consistent with the model exploiting
this, but did not rule it out.

\textbf{Run-to-run variance.} Ten runs per arm supports an exact
non-parametric test and is still a small sample. Per-run FULL suppression
counts range from 1 to 10; the precision distributions overlap at their
edges. We report medians with the full per-run spread rather than one
number.

\textbf{Model-blinded, not human-blinded, adjudication.} The study's
strongest secondary result rests on a model adjudicator. We call this
supporting evidence, not a final human-adjudicated result, and Sec.~\ref{sec:museum}'s sixth
entry demonstrates directly why an unsupervised instrument in this
pipeline needs a second, independent check.

\textbf{Incomplete ground truth.} The 20-entry curated recall list was
frozen before the study and is not the target's full vulnerability
inventory. The pilot's cross-run findings beyond that list suggest the true
positive surface is larger than the list captures, so recall is a lower
bound against an incomplete reference, not a ceiling. The list also
double-counts one condition -- the login SQL-injection finding satisfies
both an \texttt{sqli} and an \texttt{auth\_bypass} entry at the same
endpoint, in all 20 runs -- which inflates the absolute recall figures in
Sec.~\ref{sec:study-results} by roughly five percentage points uniformly across both arms; the
deduplicated, 19-condition figures (0.211 FULL vs.\ 0.158 NO-VERIFY) are
reported alongside the raw ones for that reason.

\textbf{Unreliable coverage metrics.} A known accounting defect (Sec.~\ref{sec:museum},
Entry~5) makes this framework's own coverage percentage unreliable on this
target. We excluded it from the pre-registered endpoints for that reason
and do not report it as a result.

\textbf{Untested layers.} The scheduling and ranking-layer determinism
mechanism and the grounding-critic gate described in the source handbook
are both part of the system's design, and neither was exercised here. The
pilot's attempt to ablate the critic gate was contaminated by demand
characteristics in its own prompt (Sec.~\ref{sec:pilot}) and is not reused as evidence
anywhere in this paper.

\textbf{Scope of the claim.} This is a study of one verifier-and-acceptance
stage across two lab targets, with precision discrimination coming from one
of them. It is not a benchmark against any competing offensive-security
tool and makes no comparison of that kind.

\section{Related Work}\label{sec:related}

We position this paper as a control-and-reporting-contracts contribution: a
measurement of what one verifier-and-acceptance stage does to an offensive
agent's reported output, not a claim of superior autonomous exploitation.

\citet{anthropic2024buildingeffectiveagents} distinguishes predefined
code-path workflows from model-directed agents and describes the cost and
latency tradeoffs of adding programmatic gates around a model. This paper
turns one such gate from a design recommendation into a pre-registered,
exact-statistics ablation, with a disclosed failed pilot alongside the
result that survived it. \citet{owasp2025excessiveagency} recommends
minimal tool functionality, downstream authorization, complete mediation,
and independent approval of consequential actions for agentic systems
generally; the stage we study is one concrete instance of that last
recommendation, measured rather than only argued for.

\citet{deng2023pentestgpt} studies LLM-assisted penetration testing
directly and documents the difficulty of maintaining an integrated
understanding of a testing task across a long, model-driven engagement. We
do not claim better autonomous exploitation than that or any other system;
our two arms, in fact, repeatedly found several of the same core
vulnerability classes (Sec.~\ref{sec:study-results}), with substantial per-run variation --
this paper isolates a control layer's effect on what ships, not a claim of
identical exploit-finding capability across arms.
\citet{nakano2025structuredattacktrees} constrains an offensive agent's
reasoning with a deterministic, ATT\&CK-based task tree and evaluates task
completion and query counts, overlapping our broad approach of using
deterministic structure to keep an offensive agent productive and legible.
Our verifier-and-acceptance stage sits at a different pipeline point -- after
a finding is proposed, at acceptance and reporting time -- and our endpoint
is what ships and how precise it is, not how efficiently the model reaches
a task. \citet{anthropic2025sandboxing} describes filesystem and network
isolation as containment for an autonomous coding agent; that is transport-
and process-level containment, complementary to but distinct from the
application-level verifier-and-acceptance and severity-integrity stage
measured here, which operates after a request has already been made rather
than by preventing the request itself.

None of these comparisons claims the cited systems lack a comparable stage,
and none is refuted by anything here. Our contribution is narrower and more
specific: a small, pre-registered, exact-statistics measurement of removing
one verifier-and-acceptance stage from one offensive-agent pipeline, plus a
disclosed account of the mechanism defects the act of measuring it
surfaced.

\section{Conclusion and Reproducibility}\label{sec:conclusion}

Keeping a model probabilistic and making the layer around it deterministic
is a design choice, not a discovery; the discovery reported here is
narrower and more specific -- what happens, measured, when one such layer
is switched off. Removing a verifier-and-acceptance stage eliminated
pre-report noise suppression and reduced label-blinded shipped precision,
with no statistically significant recall difference (two-sided $p=0.158$;
equivalence not established), on one public lab target, across
model-family orchestrators, under a design frozen before the first run.
Both arms repeatedly found several of the same core vulnerability classes,
with substantial per-run variation; the stage's clearest measured effect is
on what got reported, not on what got found. The factorial then attributed
the effect: the model verifier reproduces it alone; the deterministic
acceptance rules alone marked no false positives in 40 runs and did not move
model-blinded precision, re-scoping their defensible role to severity governance,
duplicate control, and auditability; and the full design's measured
model-adjudicated sensitivity point estimate was 0.938, but its lower
one-sided 95\% bound (0.875) did not clear the pre-registered 0.90 floor, a
cost we report at the same prominence as the benefit. We report every effect as a tested
median difference with its overlap stated beside it, because the data does
not support a stronger claim and a stronger claim would not survive the
scrutiny we applied to our own pilot.

The frozen pre-registrations, aggregate result tables, analysis scripts,
deviation logs, and blank human-adjudication templates are retained with the
study. Raw per-run captures and withheld arm keys are not part of this paper
release because they contain implementation detail and sensitive lab-only
evidence. A separately sanitized artifact release can provide the remaining
reproducibility materials without exposing those records. The single most
important next step is running the human blind passes against the retained,
label-stripped packets, designed as drop-in replacements for the model-blinded
passes reported in Sec.~\ref{sec:study} and Sec.~\ref{sec:factorial}. Until
those passes exist, every adjudicated result stands as supporting evidence,
not a closed claim. The
grounding-critic gate and the ranking-layer determinism mechanism this
system also implements remain entirely untested and are not claimed here
in any form.

An open companion handbook and offline harness documents the
concept-level control layers of Sec.~\ref{sec:system} in a clean-room, non-proprietary
reference implementation, so the mechanisms studied here can be built,
read, and tested independently of the implementation evaluated
here.\footnote{\url{https://github.com/mouteee/autonomous-offensive-llm-handbook}. See also
\citet{companionhandbook2026}.} The closing honesty this paper asks to be
judged on is not only its $p$-values: our own measurement pipeline broke
once, was caught by the same reconciliation discipline this system applies
to the findings it produces, and is reported here in full rather than
fixed quietly and left out.

\bibliographystyle{plainnat}
\bibliography{refs}

\end{document}